\documentclass{iau} 
\usepackage{graphicx}

\title[Light curve analysis of Cepheid Variables] 
{Multiwavelength light curve analysis of Cepheid variables}

\author[Bhardwaj et al.]   
{A. Bhardwaj$^{1}$, 
          S. M. Kanbur$^{2}$,
           M. Marconi$^{3}$,
           H. P. Singh$^{1}$,\\
           M. Rejkuba$^{4}$
\and           C.-C. Ngeow$^{5}$} 

\affiliation{$^1$Department of Physics \& Astrophysics, University of Delhi, Delhi 110007, India.\\email: {\tt anupam.bhardwajj@gmail.com} \\[\affilskip]
                $^2$State University of New York, Oswego, New York 13126, USA.\\[\affilskip]
	        $^3$INAF-Osservatorio astronomico di Capodimonte, Via Moiariello 16, 80131 Napoli, Italy.\\[\affilskip]
                $^4$European Southern Observatory, Karl-Schwarzschild-Stra\ss e 2, 85748, Garching, Germany.\\[\affilskip]
	        $^5$Graduate Institute of Astronomy, National Central University, Jhongli 32001, Taiwan.\\[\affilskip]
}

\pubyear{2018}
\volume{IAUS 339}  
\jname{Southern Horizons in Time-Domain Astronomy}
\editors{R.E.M. Griffin}
\begin{document}

\maketitle

\begin{abstract}
We present results from a detailed analysis of theoretical and observed light curves of classical Cepheid variables
in the Galaxy and the Magellanic Clouds. Theoretical light curves of Cepheid variables are based on non-linear convective 
hydrodynamical pulsation models and the observational data are taken from the ongoing wide-field variability surveys. The 
variation in theoretical and observed light curve parameters as a function of period, wavelength and metallicity
is used to constrain the input physics to the pulsation models, such as the mass-luminosity relations obeyed by Cepheid variables. 
We also account for the variation in the convective efficiency as input to the stellar pulsation models and its impact on the theoretical 
amplitudes and Period-Luminosity relations for Cepheid variables.

\end{abstract}

\firstsection 
\section{Introduction}
Classical Cepheid variables are well-known standard candles and fundamental tracers of young stellar populations in their host galaxy. Cepheid variables exhibit
a very strong Period-Luminosity relation (P-L, \cite[Leavitt \& Pickering 1912]{leavitt1912}) that has been used extensively for extragalactic distance
determination and to estimate an accurate and precise value of the Hubble constant (\cite[Freedman et al. 2001]{freedman2001}, \cite[Riess et al. 2016]{riess2016}). 
These radially pulsating stars are also very sensitive probes for the theory of stellar evolution and pulsation (\cite[Cox 1980]{cox1980a}).

Theoretical studies based on non-linear, convective hydrodynamical pulsation models by \cite{bono1999b, bono2000, marconi2005}, 
\cite[Marconi et al.(2013, and references within)]{marconi2013} have been able to predict the observed pulsation properties and the morphology 
of the light curves of Cepheid variables. In terms of mean-light properties, theoretical P-L relations for Cepheid variables were found 
to be consistent with observations at multiple wavelengths by \cite{caputo2000b}, \cite{fiorentino2007} and \cite{bono2010}. More recently, 
\cite{marconi2017} used pulsation models to match the observed light and radial velocity variations of fundamental and first-overtone mode 
Cepheids in the Small Magellanic Cloud (SMC) at multiple wavelengths. Similar efforts to reproduce Cepheid properties are made using 
the stellar evolution models, for example, \cite{anderson2016} investigated the effect of rotation and showed that
Cepheid luminosity increases between the crossings of the instability strip. Despite the recent
progress, there are several challenges for stellar pulsation modelling, such as, reproducing light variations close to the red-edge of the instability strip, disentangling the effect of helium and metallicity dependence on Cepheid properties \cite[(see, Marconi 2017, for a detailed review)]{marconi2017a}.

Over the past few years, a huge amount of variable star data became available at multiple wavelengths from the time-resolved wide-field variability surveys. 
\cite{bhardwaj2015} exploited time-series data from these large surveys to analyse light curves of Cepheid variables in the Galaxy and the Large Magellanic 
Cloud (LMC) at optical, near-infrared and mid-infrared wavelengths. \cite{bhardwaj2017} extended this work to carry out a comparative study of theoretical and observed light curve parameters of Cepheid variables at multiple wavelengths to explore constraints for stellar pulsation models. We summarize the main results from these analyses for classical Cepheids in the following sections.

\section{Theoretical and observational framework}

We use full amplitude, non-linear, convective hydrodynamical models to generate Cepheid light curves as discussed in \cite{marconi2013}.  
In brief, for a fixed composition representative of Cepheid variables in the Galaxy and the Magellanic Clouds, we adopt mass-luminosity (M-L) relations 
predicted from stellar evolutionary calculations - Canonical relations. For a given mass, we also adopt a luminosity level brighter by 0.25 dex 
to account for a possible mass-loss and overshooting - Non-canonical relations. We explore a wide range of temperatures for each combination 
of X,Y,Z and M-L to produce bolometric light variations. These bolometric light curves are transformed into visual and near-infrared 
filters. The corresponding observational dataset consists of the observed light curves of Cepheid variables at multiple wavelengths, compiled from the literature 
in \cite{bhardwaj2015}. The light curves of Cepheid variables can be analysed using the Fourier decomposition method as suggested by \cite{slee1981}. 

We fit a Fourier sine series to the light curves of Cepheid variables in the following form:
$m = m_{0}+\sum_{k=1}^{N}A_{k} \sin(2 \pi k x + \phi_{k}),$
\noindent where $m$ represents the observed magnitude and $x$ is the pulsation phase. The optimum order-of-fit ($N$) is determined based on the size of
least-square residuals. Fourier amplitude ratios and phase differences are defined as: 
$R_{k1} = \frac{A_{k}}{A_{1}} ;~ \phi_{k1} = \phi_{k} - k\phi_{1}, \mathrm{for}~ k > 1.$
\cite{slee1981} suggested that the lower-order Fourier parameters are sufficient to reproduce most characteristic features of the light curves of Cepheid
variables.

\section{Comparison of light curve parameters of Cepheid variables}

\begin{figure*}
\includegraphics[width=1.0\textwidth]{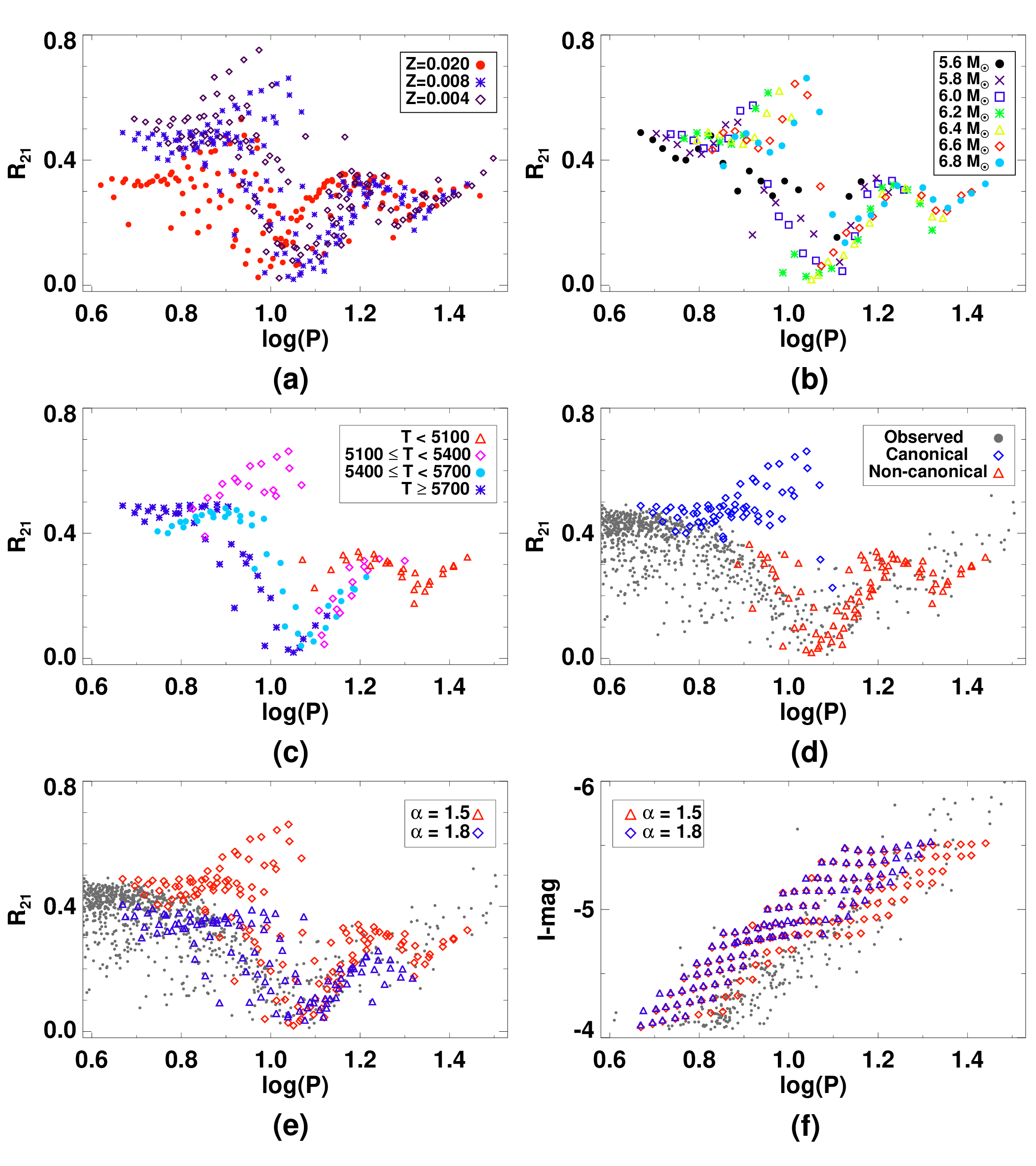}
\caption{Variation of $I$-band amplitude ratio ($R_{21}$) with period for Cepheids in the LMC as a function of: (a) metal-abundance, (b) stellar mass, (c) temperature, (d) luminosity (canonical models are fainter by 0.25 dex than non-canonical models) and (e) mixing-length ($\alpha$). Models in panels (b)-(f) correspond to the composition, Y=0.25, Z=0.008, representative of Cepheids in the LMC. In (d,e), we also compare the results from theoretical models with $R_{21}$ values from the observed light curves of Cepheids in the LMC. In (f), $I$-band P-L is shown for different values of $\alpha$ and compared with observed P-L relation for classical Cepheid variables in the LMC. \label{fig:fig1}}
\end{figure*}

We compare the theoretical and observed light curve parameters of Cepheid variables at multiple wavelengths and the detailed discussion on the variation of
Fourier amplitude and phase parameters with period, wavelength and metallicity can be found in \cite{bhardwaj2015} and \cite{bhardwaj2017}. Figure~\ref{fig:fig1} 
shows the variation of $I$-band Fourier amplitude ratio ($R_{21}$) with period as a function of different pulsation models' input parameters.  
In Figure~\ref{fig:fig1}(a), the variation in $R_{21}$ is shown as a function of metallicity. 
We find that $R_{21}$ values increase with decrease in metal-abundance for short-period Cepheid models ($\log(P) < 1$). We also note that the central
minimum around 10 days shifts to longer periods for lower metal-abundance. \cite{bhardwaj2015} also found that the central period of the Hertzsprung progression
\cite[(Hertzsprung 1926)]{hertzsprung1926} shifts to longer periods with decrease in metallicity and also for longer wavelengths. Figure~\ref{fig:fig1}(b) displays the variation in $R_{21}$ as a function of stellar mass and Figure~\ref{fig:fig1}(c) provides the same but as a function of temperature. Figure~\ref{fig:fig1}(d) shows 
the variation in theoretical $R_{21}$ values for canonical and non-canonical M-L relations and presents a comparison with $R_{21}$ values obtained from 
the observed light curves of Cepheid variables in the LMC from OGLE-IV catalogue (\cite[Soszy\'nski et al. 2015]{soszynski2015}). A comparison of 
plots (Figure~\ref{fig:fig1}b-d) suggests that canonical models are discrepant with respect to observations in the period range, $0.8 < \log(P) < 1.1$. 
These canonical set of models have masses greater than $6M_\odot$ and relatively lower temperatures ($5100 \leq T < 5400$ K), suggesting that these models
lie closer to the red-edge of the instability strip. \cite{fiorentino2007} have shown that an increase in the convective efficiency narrows 
the width of the instability strip as the red-edge becomes hotter. Figure~\ref{fig:fig1}(e) also shows that the discrepancy in $R_{21}$ values can 
be remedied by increasing the mixing-length parameter. This however, results in offset of the bolometric mean-magnitudes and affects the 
slope and zero-point of the theoretical calibrator P-L relations, as shown in Figure~\ref{fig:fig1}(f). The primary and secondary minimum in $R_{21}$
values around 10 and 20 days (Figure~\ref{fig:fig1}c), respectively, can also be correlated with the observed non-linearities in Cepheid P-L relations in the LMC at similar periods as found in \cite{bhardwaj2016}.

\section{Conclusions}

We presented a detailed light curve analysis of classical Cepheid variables in the Galaxy and the Magellanic Clouds and explored constraints for stellar pulsation
models. The variation of light curve parameters as a function of period, wavelength and metallicity shows that canonical and non-canonical models can be 
differentiated on the Fourier plane. At optical wavelengths, the amplitude parameters display a greater offset with respect to observations that can be resolved by 
increasing the convective efficiency in the pulsation models. A more quantitative and comparative light curve analysis can provide deeper insights into the theory of stellar evolution as the input physics to the pulsation models is dependent on the stellar evolutionary calculations. Light curve analysis also allows us to study Cepheid properties as a function of pulsation phase that are used to probe the interaction between stellar photosphere and hydrogen ionization front, for example in \cite{smk1993} and \cite{bhardwaj2014}. Further, light curve analysis is essential to construct templates for the identification and classification of variable stars in the era of upcoming time-domain surveys.


\end{document}